\documentclass[12pt]{iopart}
\usepackage[dvips]{epsfig}
\newcommand{\degre}{ ^{\textrm o}}

\begin{document}

\title[A new capacitive sensor]{A new capacitive sensor for displacement
measurement in a surface force apparatus}

\author{F. Restagno\dag \footnote[3]{To
whom correspondence should be addressed
(frederic.restagno@ens-lyon.fr)}, J. Crassous\dag, E. Charlaix
\ddag, M. Monchanin \ddag}

\address{\dag\ Laboratoire de Physique (UMR 5672), ENS Lyon, 46 all\'ee
d'Italie, 69364 Lyon Cedex 07
(France)}
\address{\ddag D\'epartement de Physique des Mat\'eriaux (UMR 5586),
Universit\'e Lyon I, 43 bd du 11 Novembre 1918, 69622 Villeurbanne Cedex
(France)}

\begin{abstract}
We present a new capacitive sensor for displacement measurement in
a Surface Forces Apparatus (SFA) which allows dynamical
measurements in the range of $0-100$~Hz. This sensor measures the
relative displacement between two macroscopic opaque surfaces over
periods of time ranging from milliseconds to in principle an
indefinite period, at a very low price and down to atomic
resolution. It consists of a plane capacitor, a high frequency
oscillator, and a high sensitivity frequency to voltage
conversion. We use this sensor to study the nanorheological
properties of dodecane confined between glass surfaces.

\end{abstract}

\pacs{07.05.Fb,07.10.Cm,68.35.Gy,68.45.Gd,83.85.-c,83.85.Jn}

\smallskip

{\bf Keywords :} surface forces apparatus, capacitive sensor,
nanorheology


\maketitle

\section{Introduction} \label{intro}

The surface-force apparatus developed by Tabor and Winterton
(1969) and further refined by Israelachvili and Adams (1978),
Klein (1983) and Parker {\it et al.} (1989,) has proven useful for
the measurement of colloidal forces between atomically smooth
transparent surfaces in liquid and gases at molecular scale. In
these classical apparatii the distance between the surfaces is
measured by the interferometry of white light fringes (fringes of
equal chromatic order, FECO). This technique allows to measure
steady or slowly varying distances, with a resolution of a few
$0.1 \textrm{ nm}$. Chan \etal (1985) has employed videocameras to
record rapidly changing position of the surfaces with a time
resolution about $0.5 \textrm{ s}$, during the drainage of a fluid
out of the contact region. Recently, this interferometric method
has been improved (Grunewaled (1996)) by using expensive high
speed video treatments. The first method proposed for doing a
dynamical measurement was to use a piezoelectric bimorph (Van
Alsten (1988), Israelachvili (1989), Peachey (1991)), which has
also the advantage of allowing opaque surfaces to be used.
Although Parker (1992) showed that a bimorph can be used to take a
measurement from less than a tenth of a second to several minutes,
these devices are unsuitable for measurements that take place over
many minutes or hours. Furthermore, the single-cantilever
construction of the bimorph implies that a displacement of his
head results also in an angular rotation. The resulting shear
motion makes it unsuitable for the measurement of adhesive forces.
Recently the use of a capacitor dilatometry attachment for the
conventional surface-force apparatus has been proposed by Stewart
(1992) for statics measurements with a resolution of $0.1 \textrm{
nm}$. A. Tonck {\it et al.} (1988, 1989) described a surface
forces apparatus in which they used capacitors to obtain both the
distance and the interaction force between a sphere and a plane.
This apparatus is suitable for non tranparent surfaces and
dynamical study of confined liquids.

    In this paper, we propose a new method for measuring displacement
at the nanometer scale in Surface Force Apparatus, based on
a capacitor included in an oscillator. Unlike the capacitive
sensor proposed by Steward (2000) or Franz et al. (1996,1997),
our method for the capacitance measurement is a low cost method
which does not require the use of a lock-in amplifier,
without loss in resolution or dynamic performances. From
the point of view of surface forces measurements, the method
has the advantages of being linear on a large scale,
of allowing mesasurements between non transparent surfaces,
and is suitable for dynamical measurements.
 When used in conjunction with an interferometric
technique (Crassous {\it in
preparation}) for the purpose of calibration, this sensor can be
used to perform nanorheological measurements or contact forces
measurements between the surfaces.

\section{The device}

\subsection{The surface force apparatus}

A schematic diagram of the surface forces apparatus is given in
figure \ref{fig:sfa_plan}. This apparatus has several features
which distinguish it from the common SFA. First of all, the
surfaces are not necessary transparent, since the SFA does not use
the FECO technique. The surfaces are usually a sphere and a plane.
The plane surface is mounted on the left-hand double-cantilever
$L_1$ of stiffness $2950 \textrm{ Nm}^{-1}$. An optical
interferometer measures the deflection of $L_1$ to obtain directly
the force measurements. The sphere is mounted on the right-hand
double-cantilever $L_2$ and can be moved in the direction normal
to the plane. The cantilever $L_2$ prevents the rolling of the
surfaces. The sphere motion is controlled by three actuactors. The
first one is a motorized microscrew driven by a stepping motor. It
allows a displacement of $30$~nm to $5$~cm  and is used for a
rough positionning of the sphere. The second actuator is a
piezoelectric actuator which allows a continuous approach of the
two surfaces with a velocity range of $0.1$ to $100$~nm.s$^{-1}$.
The last piezoelectric actuator is designed to add a small
sinusoidal motion in to study the dynamic behavior of the
sphere-plane interactions. The relative displacement between the
sphere and the plane, $h$, is determined by the capacitive sensor
described in this article. Finally, in order to calibrate the
capacitive sensor, a permanent magnet mounted on the cantilever
$L_1$ is located in the magnetic field gradient produced by a
little coil of copper wire. This  setup allows to calibrate the
sensor over a large range of displacement ($1$~$\mu$m).

All these devices are controlled by a Hewlett Packard VXI 743
computer equiped with a E1421A 16 channels A/D and D/A converters.

A more complete description of this apparatus will be given in a
forthcoming publication.

\subsection{The capacitor sensor}

 The measurement capacitor consists
of two duraluminium discs with a radius $R=3\times 10^{-2}
\textrm{ m}$ and a thickness $1 \textrm{ mm}$. The typical
distance between the plates is typically $d=90$~$\mu\textrm{m}$
and the surfaces have been polished to have a roughness smaller
than the sistance between the two plates of the capacitor. One
plate is fixed on the cantilever supporting the plane $L_1$, and
the other on the cantilever supporting the sphere $L_2$ so that
when the surfaces are brought together, the plates of the
capacitor do also. Parallel alignment of the plates is obtained
with a mechanical ball-and-socket joint which is rigidly screwed
after the plates have been pushed in contact to obtain the
parallelism. The terminals of the capacitor plates are connected
to the oscillator with thin copper leads whose compliance is much
higher than the one of the cantilever. To decrease the viscous
drag induced by the air flow between the plates of the capacitor,
some holes are drilled in the moving plate. The weight of the
capacitor, which is important for the resonant frequency of our
surface force apparatus is roughly $m \sim 12 \textrm{ g}$.

The capacitance $C$ of this sensor is typically $300$~pF and its
serial resistance about $1~\Omega$.

In order to measure the capacitance variations of this sensor, we
include it in an oscillator. We use a Clapp oscillator containing
two fixed capacitor $C_1,C_2$, the variable capacitor $C$, and an
inductance $L$ (figure \ref{fig:oscillateur}). The Clapp
oscillator is known to have a good stability and to be easy to
build (Audouin \etal (1991)). Neglecting the leads capacitances
and straight capacitances, the frequency of the oscillations of
the Clapp oscillator is~:
\begin{equation}\label{equ:freq}
f=\frac{1}{2\pi\sqrt{L\frac{1}{1/C_1+1/C_2+1/C}}}
\end{equation}

This formula can be used to have an estimation of the nominal
frequency and of the sensitivity of the sensor. Using the typical
values $C=2.5\times 10^{-14}/(d+h)$, $C_1=C_2=220$~pF and
$L=2.2~\mu$H, we have a typical frequency of $12$~MHz and a
typical sensitivity of the order of $1$~Hz/\AA. This shows that in
order to have a precision of $0.05$~nm on the displacement
measurement, we have to read frequency variations of $0.5$~Hz. For
this purpose, we use a Hewlett Packard HP53132A counter which
reaches this precision with an acquisition time of less than
$0.1$~s. We emphasize that reading one part in $10^7$ is really
easy in a frequency measurement but is difficult and expensive in
voltage measurements.

\subsection{Static performances}

\subsubsection{Linearity} \label{sec:linearity}

The conventional way to make a distance measurement with this
device is first to fix the distance between the capacitor's plates
at a distance comprised between $50 \mu\textrm{m}$ and $100
\mu\textrm{m}$ and to calibrate the sensitivity for small
displacements around this distance. Indeed we do not use equation
(\ref{equ:freq}) to determine the sensitivity of the sensor, since
this latter depends slightly on the angular parallelism  of the
capacitors plates. In order to calibrate the sensor, we use an
interferometer, which is mounted on our SFA (Schonenberger (1989))
and allows to perform easy calibrations. The detailed procedure is
as follows: the Rhs cantilever $L_2$ is fixed, a force is applied
on the cantilever $L_1$ by the mean of the coil/magnet system. The
deflection of $L_1$ results in a displacement $x$ of the sensor's
plate. fixed on $L_1$, as well as of the mirror. The
interferometer gives access to the absolute value of $x$, and the
calibration is done by plotting $f(x)$, the frequency of the
oscillator as a function of $x$ . In order to reduce the noise
(see paragraph \ref{sec:noise}), we usually integrate the
frequency signal over a time of $1$~s. Fig. \ref{fig:f_x} shows
the calibration over a displacement range of $80$~nm. One can see
that the capacitive sensor is very linear. The typical maximum
deviation to linearity over this scale is lower than $1\%$ of the
total excursion range. The measured sensitivity
is:~7.70~Hz.\AA$^{-1}$. This is closed to the estimated value
deduce from equation (\ref{equ:freq}) but take into account all
the straight capacitances.

\subsubsection{Influence of stray capacitance}\label{sec:stray}

The capacitance measuring circuit is in fact sensitive to stray
capacitance between the upper sensing electrode and ground.
Therefore the value $C$ of the capacitance in equation (1)
includes not only the sensor capacitance, but also the value
of stray capacitances, the larger of which is the capacitance
of the screen cable connecting the sensor to the circuit.
The order of magnitude of those stray capacitances can be estimated
by increasing the distance between the capacitor electrodes
up to the point where it does not affect anymore the frequency
of the oscillator. The overall value of the stray capacitance
can be as large as $50 pF$.

During the typical time of an experiment in a SFA (typically 30 mn)
and with the environment  conditions required by the SFA itself
(the SFA is located in a separated closed room where nobody
enters during an experiment ; signal acquisition and experiment
control are performed from another room), it turns out that
the overall stray capacitance does not change significantly
except for smooth drifts which cannot be distinguished from
the thermal drift of the measuring circuit itself (see hereafter).
Significant change of the stray capacitance occur usually other
large period of time (one day) or when a change is made on the
sensor (tuning of the distance or orientation of the electrodes,
change in the location of the oscillator). Therefore, the sensitivity
of the sensor is periodically calibrated with the interferometer,
in order to take in account the changes in sensitivity induced
by the modification of the value of the stray capacitance.

\subsubsection{Noise and drift}\label{sec:noise}

Without any displacement imposed on the cantilevers, we can
measure the noise and the thermal drift in a typical situation.
Those quantities will limit the static performance of our
apparatus and the thermal drift must be corrected to obtain an
accurate measurement of the relative displacement of the surfaces.
Figure \ref{fig:vibration_noise_freq} shows a typical record of
the signal given by the counter converted in displacement. The
noise is less $0.1$~nm peak to peak. This is due to the mechanical
vibrations on the cantilever $L_1$. With a simple plexiglass cover
over the entire instrument and without any temperature control we
find a drift rate smaller than $0.01$~nm/s. This drift is of the
same order as the drift reported in other articles (Schonenberger
(1989)).

It is worthwhile to inquire about the electrostatic attractive
forces between the charged plates of this plane capacitor. In
general the force is given by:
\begin{equation}\label{equ:force_capa}
F=-\epsilon_0<V^2>S/d^2
\end{equation}
where $<V^2>$ is the average of the square voltage between the
capacitor's plates, $S$ the plates area and $\epsilon_0$ the
dielectric permittivity of vacuum. The force is $\approx
1.38$~$\mu$N for typical values $d=90$~$\mu$m and $<V>=3.5$~V.
This force is nearly constant over one experiment since the
relative displacement of the plates is always much smaller than
$d$.

\subsection{Dynamical measurements}

A piezoelectric crystal is used to add a sinusoidal motion (Tonck
(1988)) of small amplitude on cantilever $L_2$ in order to
determine the dynamic behaviour of the sphere-plane interaction.
The distance between the surfaces is thus $h$ with $h$ being the
sum of two components:
\begin{equation}
h=h_{dc}(t)+h_{ac}\cos(i\omega t)
\end{equation}
where $h_{dc}(t)$ is a slowly varying function of time ($0.01<
\dot{h}_{dc}<100$~nm/s). This results in a modulation of the
frequency of the capacitive sensor. This harmonic frequency
modulation cannot be read with the counter when $\omega/2\pi$ is
larger than $1$~Hz. We built a high-resolution frequency to
voltage converter to read this distance modulation between the
surfaces. The diagram of this converter is drawn on figure
\ref{fig:pll}. The principle of the operation is as follows : the
high frequency signal (frequency $f$) is multiplied by a reference
signal generated by a stable function generator HP31320A
(frequency $f_{ref}$). The output signal is a combination of
signals at $f-f_{ref}$ and higher frequency signals. It is first
passed through a low-pass filter then directed to a frequency to
voltage converter built with a digital phase-lock-loop with  a
range of $5.10^3$~Hz and a sensitivity of $5.10^{-4}$~V.Hz$^{-1}$.
This frequency-shift technique allows to obtain a high sensitivity
in the conversion which could not be directly obtained with a
phase lock loop. The final sensitivity on the $ac$ displacement is
$5\times 10^{-3}$~V.nm$^{-1}$.

The output voltage is connected on a digital two-phase lock-in
amplifier (Standford Research System SR830 DSP Lock In Amplifier)
whose reference signal is the signal used to drive the
piezoelectric element (see figure \ref{fig:ftoV_principle}).

The dynamical response of the displacement sensor can be obtained
with the same procedure as the static calibration. A white noise
excitation containing all the frequencies in the range $0-100$~Hz
is applied on the cantilever $L_1$ by the mean of the coil/magnet
system. The frequency response of the capacitive sensor is
calibrated by the frequency response of the interferometer mounted
on $L_1$, whose response is flat in amplitude and frequency. This
also allows a dynamical calibration of the displacement sensor.

The electrical noise  of the capacitive sensor converted in
distance is less than $1$~pm.Hz$^{-1/2}$ in the range 0-100~Hz,
except in the range $49-51$~Hz, where the noise of the electronics
is bigger than a few pm.Hz$^{-1/2}$. Since the dynamic experiments
are usually made at a given frequency, this frequency must be
chosen not to close to the line frequency.

The mechanical noise on the displacement sensor (figure
\ref{fig:vibration_spectrum} is due to the mechanical vibrations
on the cantilever $L_1$ and are much more important than the
electronic noise.

\section{Application to surface forces measurement}

\subsection{Experimental system}

In this experiments, we use a sphere of $2.7$~mm in diameter and a
plane made of Pyrex. The surfaces are washed in an ultrasound bath
with distilled water and a detergent soap  for more than an hour.
The surfaces are then rinsed with propanol purified at $99\%$.
Finally the surfaces are passed in a flame in order to burn out
the last amount of pollution and to flatten the surfaces. The
total roughness of this surfaces measured by an atomic force
microscope is less than $0.3$~nm rms on a 1~$\mu$m$^2$ square. The
surfaces are quickly mounted on the apparatus.

A small drop of an organic liquid : $n$-dodecane obtained from
Acros Organics, is placed between the surfaces. Dodecane is a
simple, Newtonian, non polar liquid. The length of this molecule
obtained by X-ray diffraction is tabulated as $1.74$~nm. The
liquid has a purety better than $99\%$. The viscosity of the
liquid is given as $1.35$~mPl in the handbook at $25\degre C$. The
experiments are carried out at ambient temperature, {\it i.e.}
$25\degre$C. The apparatus is placed in a plexiglass box which
reduces sound vibrations, and contains some desiccant (silicagel)
to dry the atmosphere and prevent the dissolution of water in
dodecane.

\subsection{Dynamical measurements : a surface forces apparatus used as a
nanorheometer}

The experiment starts with the surfaces being separated by a
distance of $500$~nm. A voltage increasing linearly in time is
applied to one of the piezoelectric actuator, so that the sphere
moves toward the plane at constant speed.
 In the same time, we impose a small oscillation of the sphere
$h_{ac}\cos(\omega
t)$, with $h_{ac}=0.80$~nm at a frequency $\omega/2\pi=64$~Hz. In
the lubrication approximation, the viscous force between the two
surfaces gives the well-known  expression~(Georges (1993)):
\begin{equation}
\label{damp_eq} F=\frac{6\pi\eta R^2}{h_T}\frac{{\textrm
d}h_T}{{\textrm d}t}
\end{equation}
In our experiments, the viscous force on $L_1$ is~:
\begin{equation}\label{equ:damp_eq2}
F=\frac{6\pi\eta \omega R^2}{h}h_{ac}\cos(\omega t+\pi/2)
\end{equation}
We can read the displacement $x_{ac}\cos(\omega t+\psi)$ induced
on the cantilever $L_1$ by the viscous flow between the sphere and
the plane. Since $64$~Hz, the value of the excitation frequency,
is above the resonant frequency of the cantilever $L_1$, the real
viscous force $f_{ac}\cos(\omega t+\phi)$ is obtained by
multiplying $kx_{ac}\cos(\omega t+\psi)$ by the mechanical
transfert function of the mass-cantilever system.

First of all, we find that the measured force is out of phase the
displacement excitation ($\phi=\pi/2$) which means that a purely
dissipative force is measured. On figure \ref{dodecane}, the
inverse of the damping coefficient $f_{ac}/h_{ac}$ is plotted as a
function of the distance $h$ between the surfaces. This curve
clearly shows the a good agreement of the lubrication theory for
all the distances greater than $6$~nm which corresponds to 5
molecular length of dodecane. The origin $h=0$ is obtained by the
linear extrapolation of the best linear fit of the experimental
data and the origin of the axis $h_{ac}/f_{ac}$. The slope of this
curve combined with equation (\ref{equ:damp_eq2}), gives the
viscosity of dodecane. We find $\eta=1.37\times 10^{-3}$~Pl which
can be compared to the tabulated value (see Handbook)
$\eta=1.35\times 10^{-3}$~Pl at $25\degre$C. This result agrees
well those of Tonck {\it et al.} (1989) on the same liquid.

\section{Conclusion}

We have presented a new capacitive sensor for surface forces
measurements allowing both static and dynamic measurements between
non transparent surfaces. This sensor does not not need any
lock-in amplifier for the statics measurements and has a very low
cost. In the static regime, this sensor has a sensitivity of
$0.1$~nm with an integration time of $1$~s. In the dynamic regime,
 combined with a frequency to voltage converter, this sensor has a
sensitivity
better than $1$~pm.Hz$^{-1/2}$.

We have used this sensor for surface force measurements between
pyrex surfaces separated by a small meniscus of dodecane. Away
from the contact between the surfaces, we have confirmed previous
results showing that the bulk viscosity of the liquid is not
affected by the confinement. At a distance between the surfaces
smaller than a few molecular length, the dissipation increases.

\ack We thank J.P. Zaygel for his help in electronics design, C.
Cottin-Bizone for his experimental help. We have benefited from
discussions with J.-L. Loubet and A. Tonck. We are happy to thank
J.-M. Combes for technical help. This work has been supported by
the Region Rh\^{o}nes-Alpes contact number 98B0316 and the
D\'el\'egation G\'en\'erale de l'Armement.

\section*{References}
\smallskip
\begin{harvard}
\item[] Audouin C, Bernard M Y, Besson, R, Gagnepain J J, Groslambert J,
Granveaud M,  Neau J C, Olivier M and Rutman J 1991 {\it La mesure de la
fr\'equence des oscillateurs}
Masson
\item[] Chan D Y C and Horn R G 1985 The drainage of thin liquid films
between solid surfaces {\it J. Chem. Phys.} {\bf 83}(10) 5311-24
\item[] Frantz P, Agrait N and Salmeron M 1996 Use of capacitance to
measure surface forces. 1. Measuring distance separation with enhanced
spatial and time resolution {\it Langmuir} {\bf12}
3289-94
\item[] Frantz P, Artsyukhovich A, Carpick R and Salmeron M 1997 Use of
Capacitance to Measure Surface Forces. 2. Application to the Study of
Contact Mechanics {\it Langmuir} {\bf 13} 5957-61
\item[] Georges J M, Millot S,Loubet J L and Tonck A 1993 Drainage of thin
liquid films between relatively smooth surfaces{\it J. Chem. Phys.} {\bf
98}
7345-60
\item[] Grunewald T and Helm C A 1996 Computer-controlled experiments in
the surface forces apparatus with a CCD spectrograph {\it Langmuir} {\bf
12} 3885-90
\item[] Handbook of Chemsitry and Physics 1964 49$^{th}$ Edition,
Ed. R.C. Weast,The Chemical Rubber Co., Cleveland 1964.
\item[] Israelachvili J N and G E Adams 1978 Measurements of forces
between mica surfaces in aqueous electrolyte solutions in the range 0-100
nm {\it J. Chem. Soc. Faraday Trans.} I {\bf 74}  975-1001
\item[] Israelachvili J N, Kott S J  and  Fetters L J 1989 Measurements of
dynamic interactions in thin films of polymer melts : the transition from
simple to complex behaviour {\it J. Polym. Sci.} B {\bf 27} 489-502\
\item[] Klein J 1983 Forces between mica surfaces bearing adsorbed
macromolecules in liquid media {\it J. Chem. Soc. Faraday Trans.} I {\bf
79} 99-118
\item[] Parker J L and Christenson 1989 H K A device for measuring the
force and separation between two surfaces down to molecular separation
{\it Rev. Sci. Instrum.} {\bf 60}(10)3135-9
\item[] Parker J L 1992 A novel method for measuring the force between two
surfaces in a surface force apparatus Langmuir {\bf 8} 551-6
\item[] Peachey J, Van Alsten J and Granick S 1991 Design of an apparatus
to measure the shear reponse of ultrathin films {\it Rev. Sci. Instrum.}
{\bf 62}(2) 463-73
\item[] Restagno F, Crassous J, Charlaix E and Monchanin M {\it in
preparation}.
\item[] Schonenberger C and Alvarado S F 1989 A differential
interferometer for force microscopy {\it Rev. Sci. Instrum.} {\bf 60}(10)
 3131-34
\item[] Stewart A M and Christenson H K 1990 Use of magnetic forces to
control distance in a surface force apparatus {\it Rev. Sci. Instrum.}
{\bf 1} 1301-3
\item[] Stewart A M 2000 Capacitance dilatometry attachment for a
surface-force apparatus {\it Meas. Sci. Technol.} {\bf 11} 298-304
\item[] Tabor D and Winterton R H S 1969 The direct measurement of normal
and retardated van der Waals forces {\it Proc. R. Soc.} A {\bf 312} 435-50
\item[] Tonck A, Georges J M  and Loubet J L 1988 Measurement of
intermolecular forces and the rheology of dodecane between alumina
surfaces {\it J. Colloid Interface Sci.} {\bf 126}(1)
150-63
\item[] Tonck A 1989 D\'eveloppement d'un appareil de mesure de forces de
surface et de nanorh\'eologie {\it Th\`ese de Doctorat} \'Ecole Centrale
de Lyon, Lyon
\item[] Van Alsten J and  Granick S 1988 Molecular tribometry of ultrathin
liquid films {\it Phys. Rev. Lett.} {\bf 61}
2570-73
\end{harvard}

\newpage
\pagestyle{empty}

\begin{figure}[htbp]
\caption{Horizontal section of the surface forces apparatus with
the capacitor stage. The sphere is moved horizontally by a
stepping motor allowing large displacements (a step is 30~nm and
the displacement range is larger than 1 cm). A piezoelectric
crystal controls the approach of the two surfaces at constant
velocities. A second piezoelectric crystal adds a small
oscillatory motion to the sphere. This allows to study the dynamic
behavior of the sphere-plane interactions. The deflection $x$ of
the first cantilever $C_1$ of stiffness $k$ measures the force
exerted on the plane by the sphere. (a):~schematic diagram.
(b):~mechanical diagram. } \label{fig:sfa_plan}
\end{figure}

\begin{figure}[htbp]
\caption{Diagram of the electronic oscillator.}
\label{fig:oscillateur}
\end{figure}

\begin{figure}[htbp]
\caption{$\bullet$ : Measured frequency $f \textrm{  (Hz})$ of the
oscillator as a function of the relative displacement between the
surfaces measured by an interferometric method. The displacement
is imposed by the coil/magnet system. The full line represents the
best linear fits of the datas. This shows the very good linearity
in the common measure range of a surface forces apparatus.}
\label{fig:f_x}
\end{figure}

\begin{figure}[htbp]
\caption{Measured frequency $f \textrm{  (Hz})$ of the oscillator
for a fixed capacitor value.}
\label{fig:vibration_noise_freq}
\end{figure}

\begin{figure}[htbp]
\caption{Diagram of the electronics of the frequency to voltage
conversion. The function generator used to provided a fixed
frequency oscillatory signals is a Hewlett Packard HP313120A
function generator.}
\label{fig:pll}
\end{figure}

\begin{figure}[htbp]
\caption{Schematic diagram of the dynamical measurements.}
\label{fig:ftoV_principle}
\end{figure}

\begin{figure}[htbp]
\caption{Vibration spectrum measured on the capacitive sensor. The
$50$~Hz signal is
 large and the other main peak are some vibration peaks observed in the
environment. Two important peaks have been indexed on this figure
a:~the resonant peak of the force cantilever, b:~the resonant peak
of the anti-vibration device.}
\label{fig:vibration_spectrum}
\end{figure}

\begin{figure}[htbp]
\caption{Plot of the inverse of the damping coefficient
$h_{ac}/f_{ac}$ as a function of the displacement $h_{dc}$, for
dodecane at $25 \degre \textrm{C}$. $h_0=0,3$~nm for $h_{dc}>6
$~nm. The arrows indicate the inward ($\leftarrow$) and outward
($\rightarrow$) approach : no hysteresis in the dynamical response
has been observed. The dotted line is the best linear fit of the
datas.}
\label{dodecane}
\end{figure}

\newpage
\psfig{file=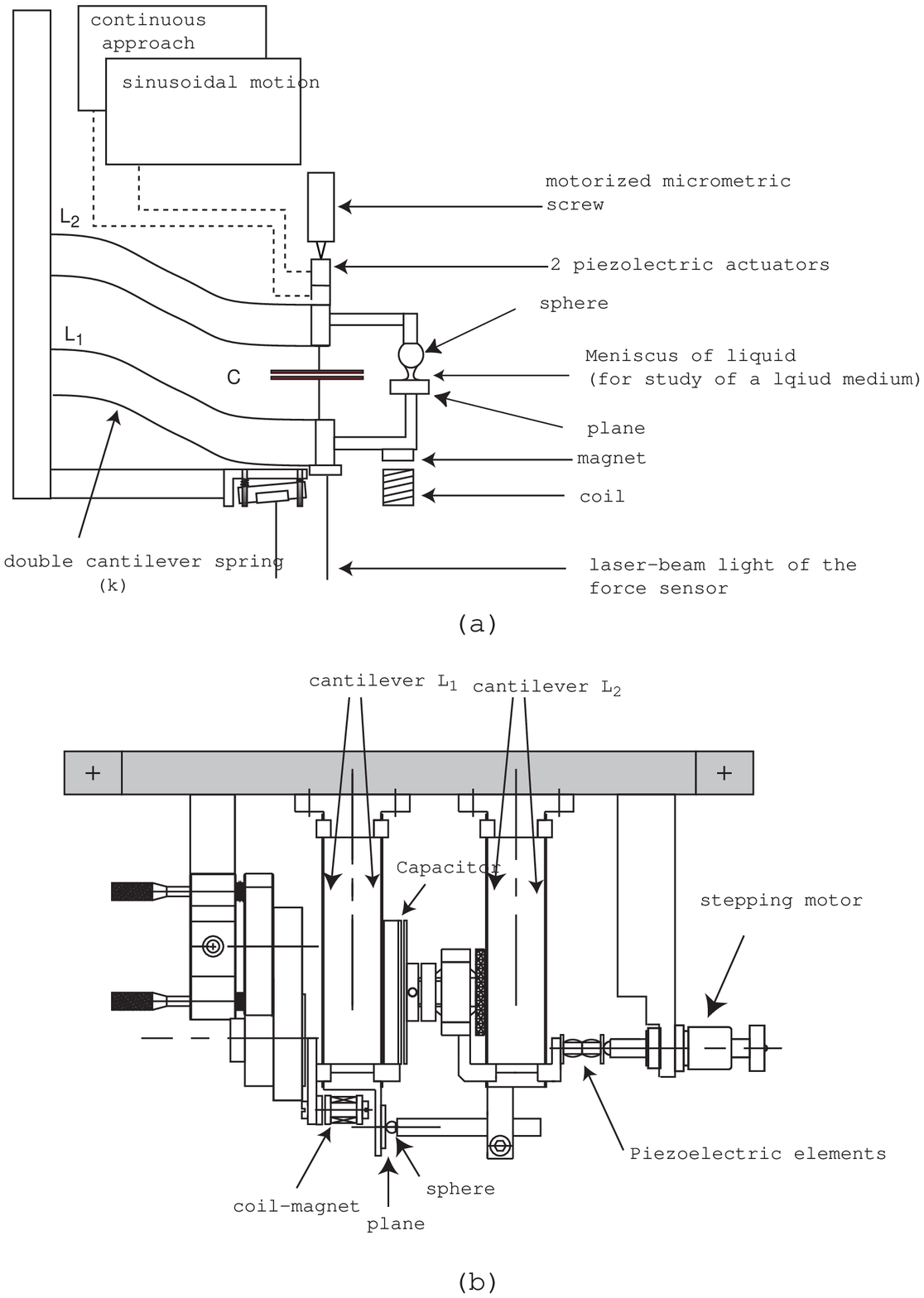,width=14cm}
\newpage
\psfig{file=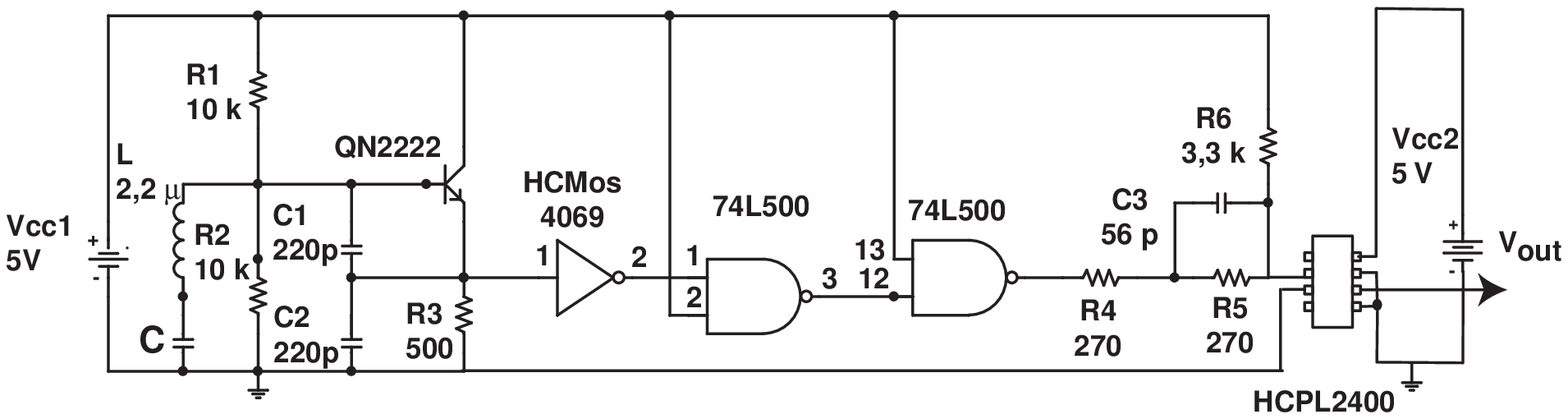,width=8.4cm}
\newpage
\psfig{file=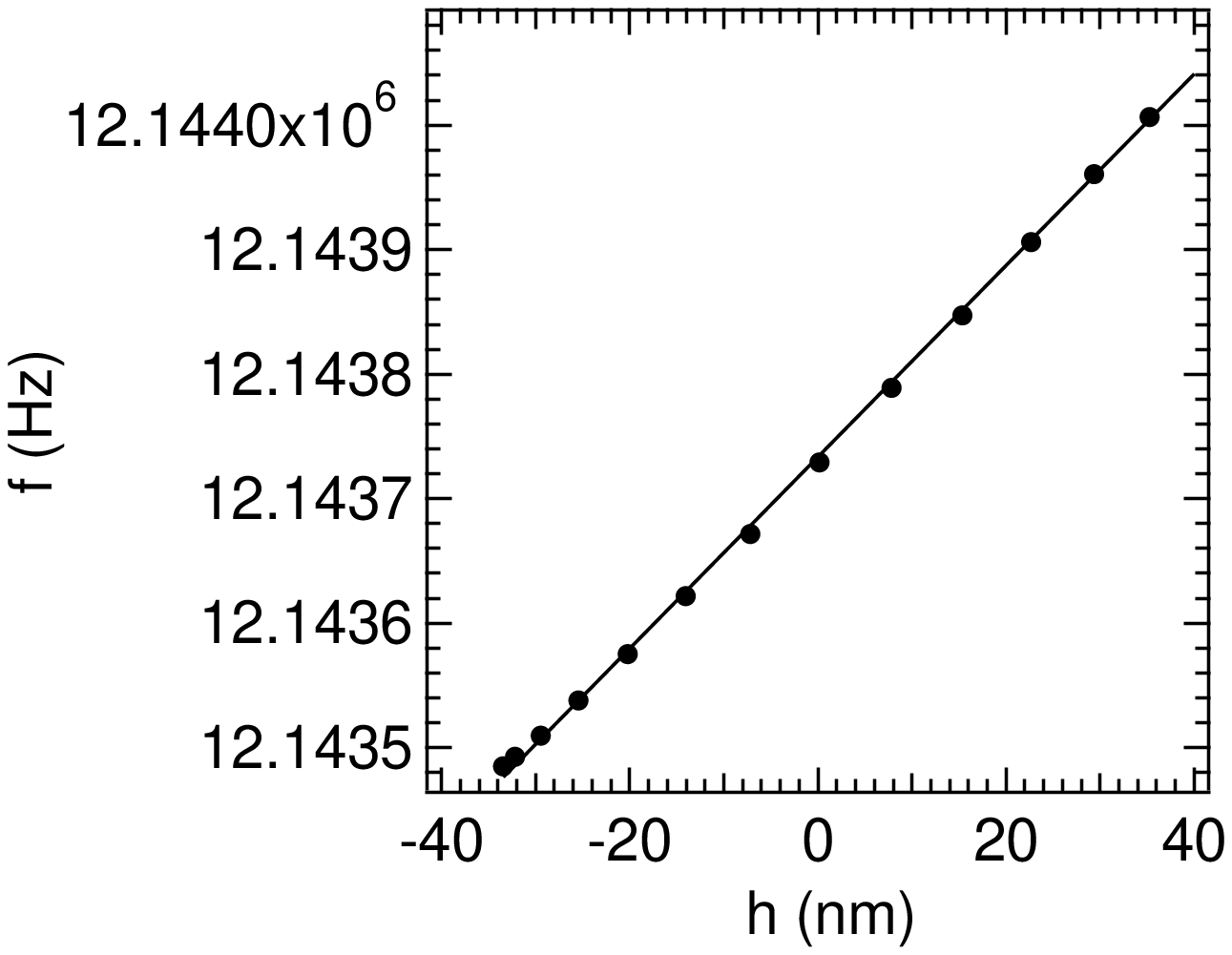,width=8.4cm}
\newpage
\psfig{file=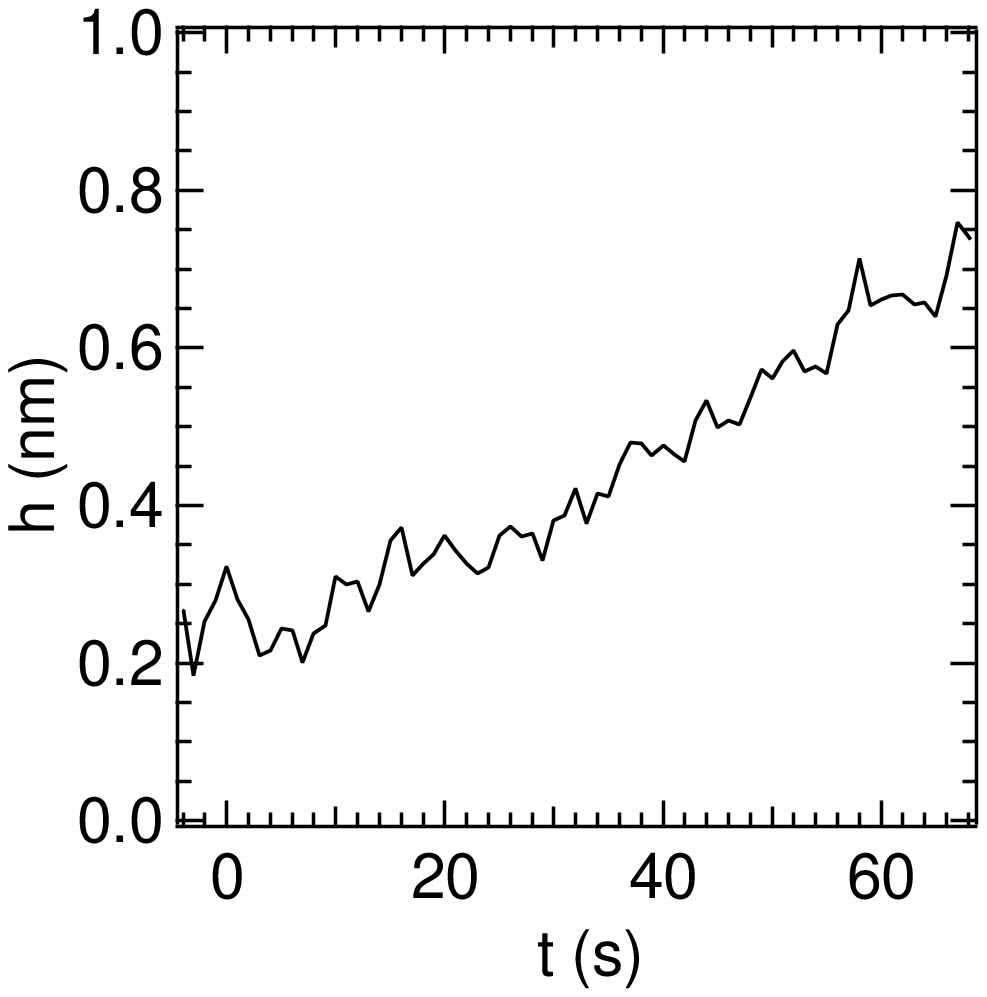,width=8.4cm}
\newpage
\psfig{file=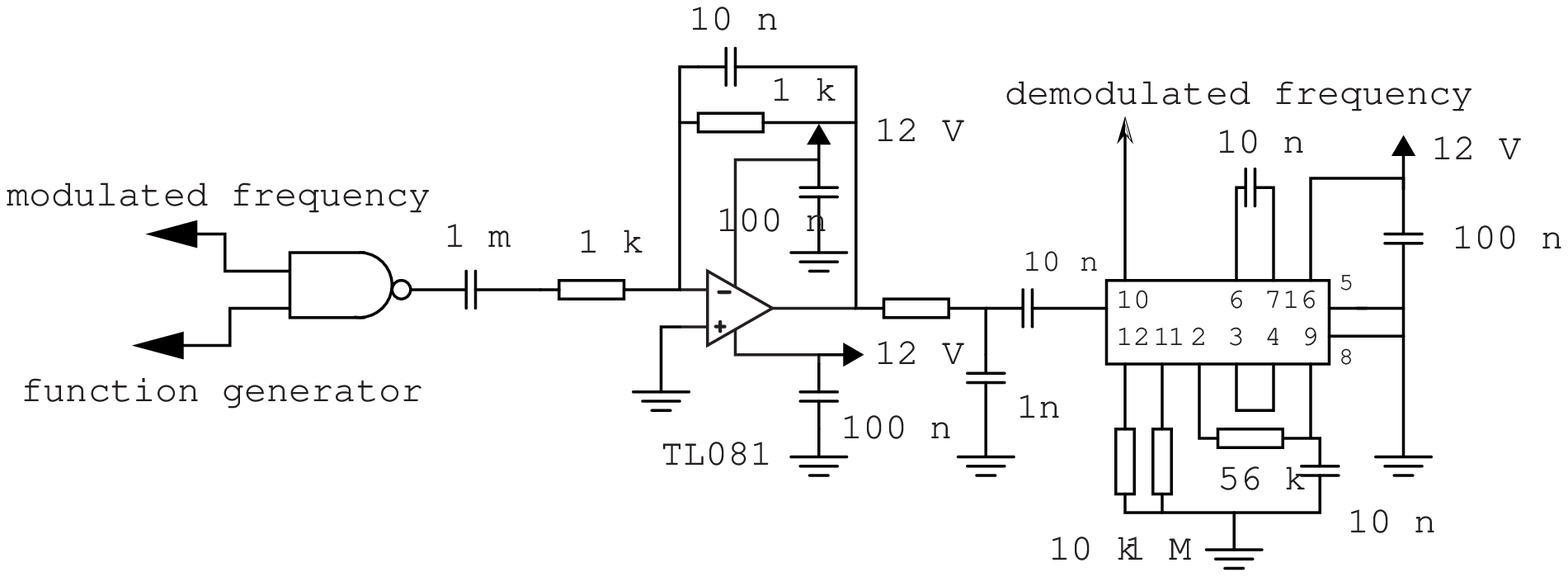,width=8.4cm}
\newpage
\psfig{file=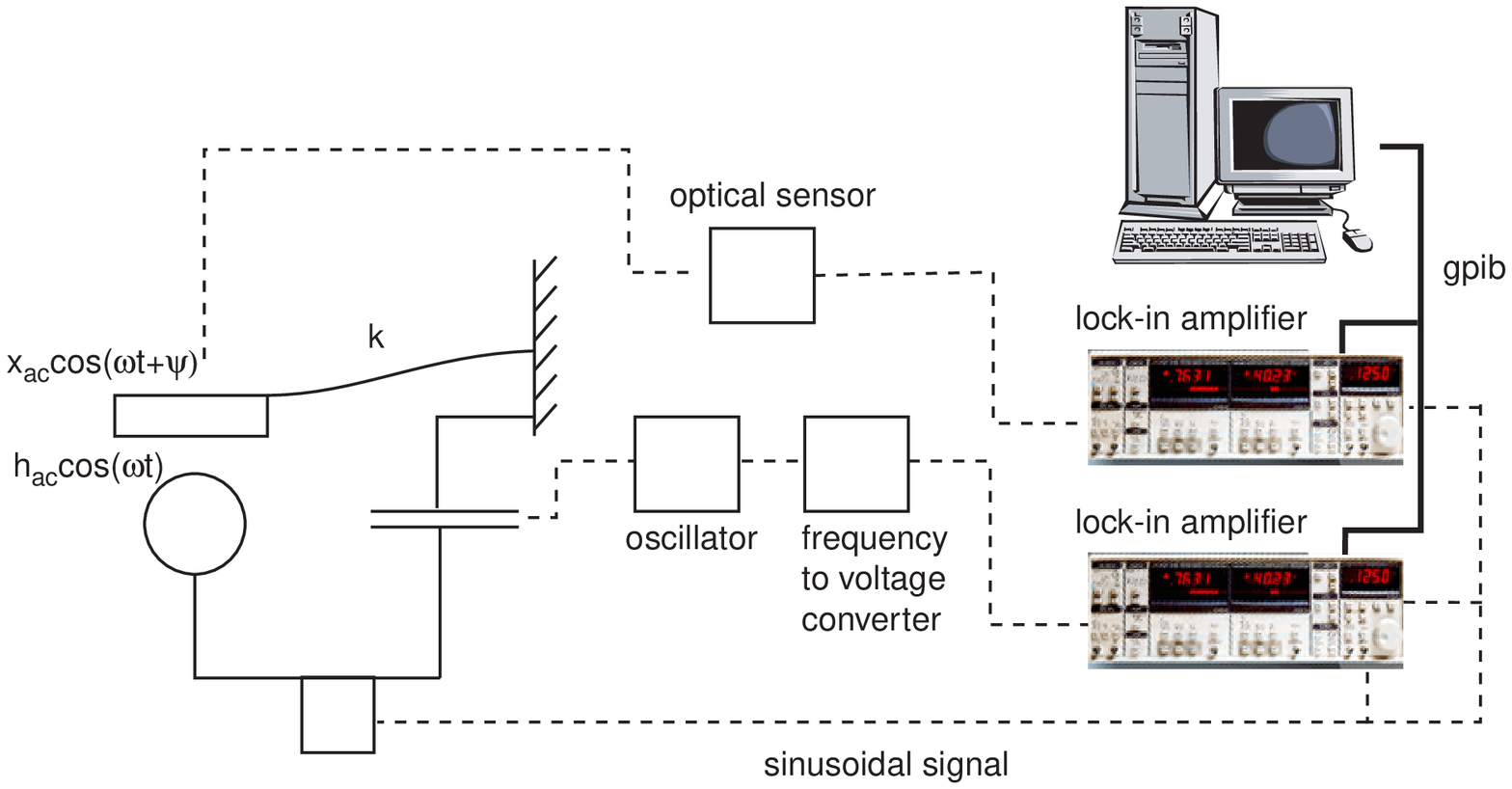,width=13cm}
\newpage
\psfig{file=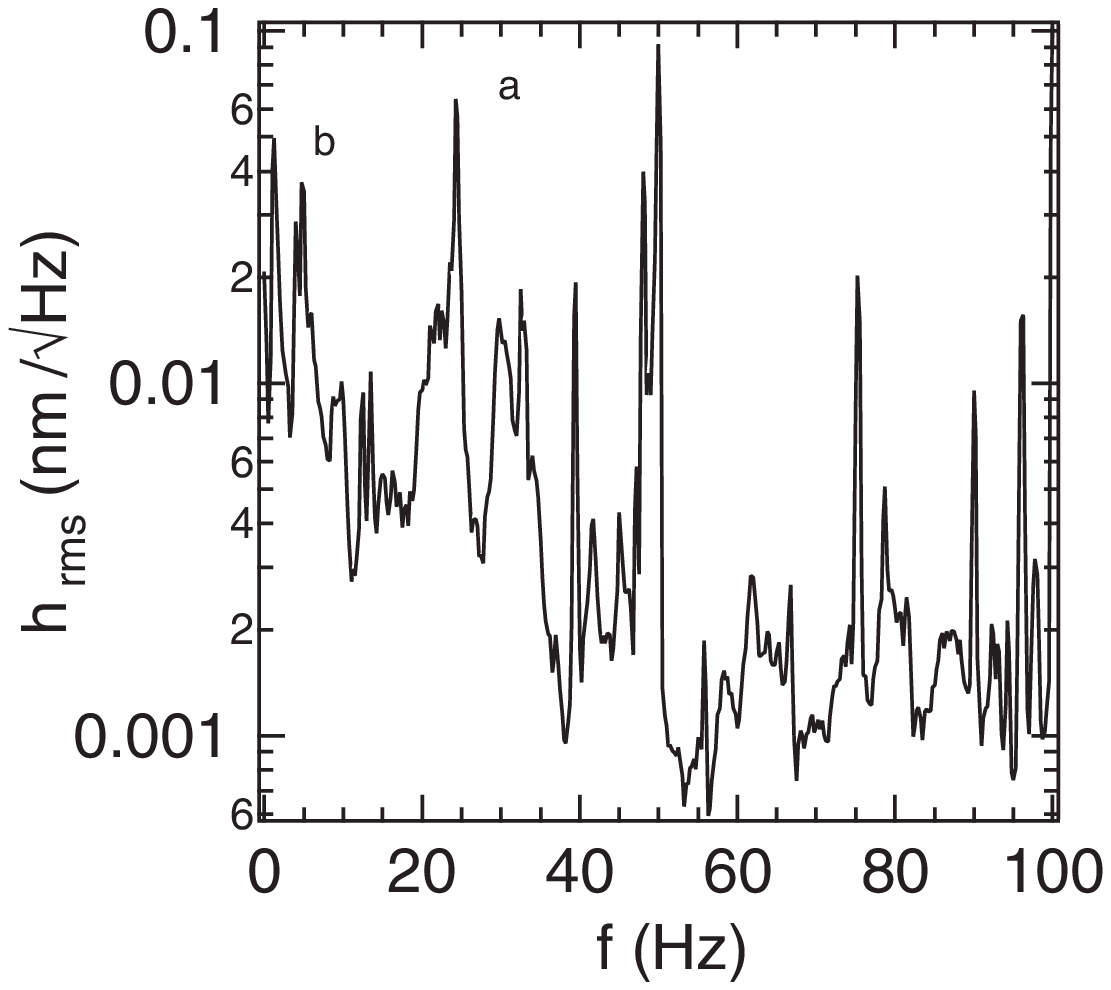,width=8.4cm}
\newpage
\psfig{file=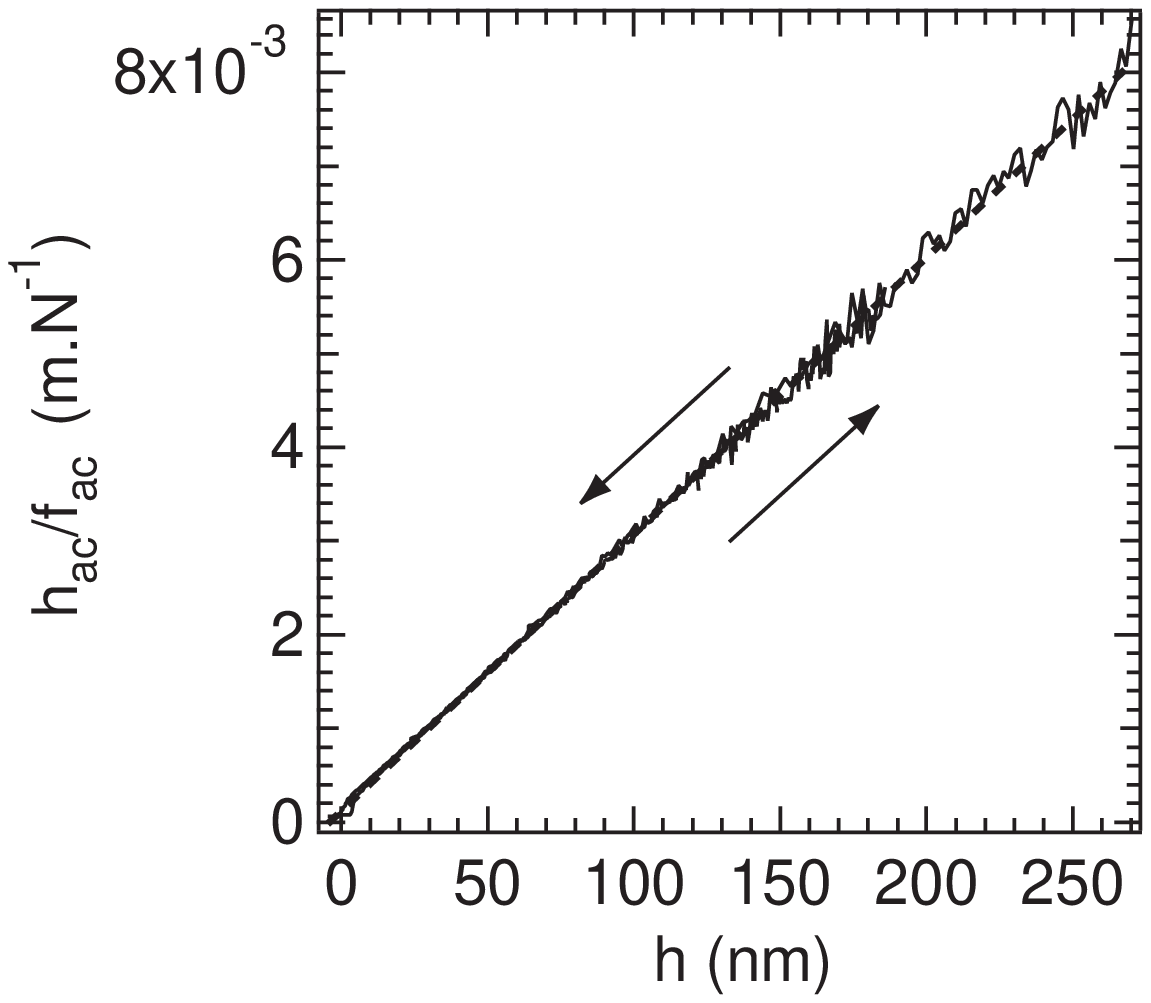,width=8.4cm}
\end{document}